# One Approach on Derivation of the Schrödinger Equation of Free Particle

Shin-ichi Inage.
*Power & Industrial Systems R & D Laboratory, Hitachi Ltd.*
*832-2, Horiguchi, Hitachi-naka-shi, Ibaraki-ken, 312-0034 Japan*

The Schrödinger equation based on the de Broglie wave is the most fundamental equation of the quantum mechanics. There can be no doubt about it's prediction validity. However, the probabilistic interpretation on the quantum mechanics has insoluble semantic interpretations like 'reduction of wave packet' on observations of physical values. Especially, it is not clear that the wave function $\Psi$ which is described by complex function, is whether '***formality***' or '***reality***' to express the state of particle motion. On this paper, we interpret the wave nature of particle as not the inherency of particle itself, but the motional property of particle in fluctuated space-time due to the kinetic energy and momentum the belief that the kinetic energy and momentum fluctuates the microscopic space-time, and the particle move through the fluctuated space-time adversely. Then, through the particle motion in the Euclidean space-time, the particle will be recognized as if it has the wave nature. We estimate the governing equation of fluctuations of microscopic space-time based on the macroscopic law of motion. On this paper, the equivalence between the governing equation and the Schrödinger equation is indicated.

## §1. Introduction

From a practical point of view, there seems little doubt that the quantum mechanics is complete. And the other, as long as the interpretation about the wave nature of the particle is uncertain, the quantum mechanics leaves a lasting impression on the woolliness as typified by the 'complementarity principle'. As well known, so many approaches were tried to understand the wave nature of particle [1]-[3]. This paper is also relevant to one approach to understand the wave nature of particle and making the connection between the standard probabilistic interpretation and the new approach. To be consistent with the quantum mechanics, the every approach assures of the derivation of the Schrödinger equation.

The particle motion is recognized and described through the Euclidean space-time on the macroscopic systems. If on the microscopic space-time the kinetic energy and momentum of particle give fluctuations to space-time and to the contrary the particle motion through the fluctuated space-time, then we may recognize the particle path with fluctuations in macroscopic space-time. As the dynamic equation is described by the macroscopic space-time system, the 'superficial' fluctuations are added to particle motions. Under such a hypothesis, using the principle of the least action due to macroscopic Lagrangian, the governing equation of the fluctuation of the microscopic space-time is estimated.

## §2. Theoretical Approach

On this paper, the free particle under the uniform motion is considered. Here, the macroscopic space-time and the microscopic space-time are expressed as (X, T), (x, t) respectively. Further, as the microscopic space-time (X, T) is the functions of $x$ and t, these functions are expressed φ and η, respectively. Then, the coordinate conversion between (X, T) and ($x$, t) are

$$dX = \frac{\partial \phi}{\partial x} dx + \frac{\partial \phi}{\partial t} dt ,\qquad(1)$$

$$dT = \frac{\partial \eta}{\partial x} dx + \frac{\partial \eta}{\partial t} dt .\qquad(2)$$

And, as dynamic law, the following hypothesizes are considered.

**Hypothesis Ⅰ:**

On the macroscopic space-time system, the speed of free particle under uniform motion is defined as follows using of the macroscopic space X and time T.

$$\frac{dX}{dT} = V \qquad(3)$$

Further, the kinetic energy E is also defined as

$$E = \frac{1}{2} m V^2 = \frac{1}{2} m \left(\frac{dX}{dT}\right)^2 .\qquad(4)$$

**Hypothesis Ⅱ:**

On the macroscopic space-time system, the particle moves along the geodesic line due to the least action. The geodesic line is defined by a function which minimize the following variation:

$$\int_0^T L\, dT = \int_0^T \frac{m}{2}\left(\frac{dX}{dT}\right)^2 dT \qquad(5)$$

The second hypothesis is the Lagrangian function L defined by (5) on the macroscopic space-time system, has the minimum value under the microscopic space-time.' i.e. the Lagrangian function L satisfies the Euler-Lagrangian equation defined by the microscopic space-time system. On the macroscopic system, (5) give the linear motion as the geodesic line of uniform motion. So, this hypothesis requires that the macroscopic geodesic line is also the one on the microscopic space-time.

First, the Hypothesis Ⅰ is considered. From (1), (2)

and (3), we have

$$\frac{dX}{dT} = \frac{\frac{\partial \phi}{\partial x}dx + \frac{\partial \phi}{\partial t}dt}{\frac{\partial \eta}{\partial x}dx + \frac{\partial \eta}{\partial t}dt} = \frac{\frac{\partial \phi}{\partial x}\frac{dx}{dt} + \frac{\partial \phi}{\partial t}}{\frac{\partial \eta}{\partial x}\frac{dx}{dt} + \frac{\partial \eta}{\partial t}} = V. \quad (6)$$

(6) leads to

$$\frac{\partial \phi}{\partial x}\frac{dx}{dt} + \frac{\partial \phi}{\partial t} = V\left(\frac{\partial \eta}{\partial x}\frac{dx}{dt} + \frac{\partial \eta}{\partial t}\right), \quad (7)$$

with help of $\dot{x}$ ($=dx/dt$) which is the particle speed defined by the microscopic space-time,

$$\dot{x}\frac{\partial \phi}{\partial x} - \dot{x}\frac{\partial (V\eta)}{\partial x} = \frac{\partial (V\eta)}{\partial t} - \frac{\partial \phi}{\partial t}. \quad (8)$$

Further, let us consider the following function $\Psi$:

$$\psi = \phi - V\eta. \quad (9)$$

Then, from (8),

$$\dot{x}\frac{\partial \psi}{\partial x} = -\frac{\partial \psi}{\partial t}. \quad (10)$$

so that means

$$\dot{x}\frac{\partial \psi}{\partial x} + \frac{\partial \psi}{\partial t} = \frac{d\psi}{dt} = 0, \quad (11)$$

hence

$$\frac{d^2\psi}{dt^2} = \dot{x}\frac{\partial}{\partial x}\left(\dot{x}\frac{\partial \psi}{\partial x} + \frac{\partial \psi}{\partial t}\right) + \frac{\partial}{\partial t}\left(\dot{x}\frac{\partial \psi}{\partial x} + \frac{\partial \psi}{\partial t}\right)$$

$$= \dot{x}^2\frac{\partial^2 \psi}{\partial x^2} + 2\dot{x}\frac{\partial^2 \psi}{\partial x\partial t} + \frac{\partial^2 \psi}{\partial t^2}$$

$$= \dot{x}^2\frac{\partial^2 \psi}{\partial x^2} - 2\dot{x}^2\frac{\partial^2 \psi}{\partial x^2} + \frac{\partial^2 \psi}{\partial t^2} = 0, \quad (12)$$

$$\therefore \frac{d^2\psi}{dt^2} = -\dot{x}^2\frac{\partial^2 \psi}{\partial x^2} + \frac{\partial^2 \psi}{\partial t^2} = 0. \quad (13)$$

This means that the function $\Psi$ which is defined in the microscopic space-time satisfies the one dimensional classical wave equation.

Next, let us consider Hypothesis Ⅱ. This is the Euler-Lagrangian equation in the microscopic space-time. The L is defined by the macroscopic space-time as follows,

$$L = \frac{m}{2}\left(\frac{dX}{dT}\right)^2 = \frac{m}{2}\left(\frac{\phi_x \dot{x} + \phi_t}{\eta_x \dot{x} + \eta_t}\right)^2 = \frac{m}{2}V^2. \quad (14)$$

Equation (14) satisfies the following Euler-Lagrangian equation,

$$\frac{d}{dt}\left(\frac{\partial L}{\partial \dot{x}}\right) - \frac{\partial L}{\partial x} = 0. \quad (15)$$

From (14), the first term in the left-hand side of (15) is

$$\frac{\partial L}{\partial \dot{x}} = \frac{\partial}{\partial \dot{x}}\left(\frac{m}{2}V^2\right) = mV\frac{\partial V}{\partial \dot{x}}. \quad (16)$$

Thus

$$\frac{d}{dt}\left(\frac{\partial L}{\partial \dot{x}}\right) = \frac{d}{dt}\left(mV\frac{\partial V}{\partial \dot{x}}\right) = mV\frac{d}{dt}\left(\frac{\partial V}{\partial \dot{x}}\right). \quad (17)$$

Further, the second term in the left-hand side of (15) is

$$\frac{\partial L}{\partial x} = \frac{\partial}{\partial x}\left(\frac{m}{2}V^2\right) = mV\frac{\partial V}{\partial x}. \quad (18)$$

Thus, (15) is

$$\frac{d}{dt}\left(\frac{\partial V}{\partial \dot{x}}\right) - \frac{\partial V}{\partial x} = 0. \quad (19)$$

where, the temporal differentiation is defined like $d/dt = \dot{x}\cdot\partial/\partial x + \partial/\partial t$. On (19), the first term of the left-hand side is estimated as,

$$\frac{\partial V}{\partial \dot{x}} = \frac{\partial}{\partial \dot{x}}\left(\frac{\phi_x \dot{x} + \phi_t}{\eta_x \dot{x} + \eta_t}\right)$$

$$= \frac{\phi_x}{\eta_x \dot{x} + \eta_t} - \frac{\eta_x(\phi_x \dot{x} + \phi_t)}{(\eta_x \dot{x} + \eta_t)^2},$$

$$= \frac{\phi_x}{\eta_x \dot{x} + \eta_t} - \frac{V\eta_x}{\eta_x \dot{x} + \eta_t} = \frac{\psi_x}{\eta_x \dot{x} + \eta_t}. \quad (20)$$

Thus, the first term of the left-hand side of (19) is

$$\frac{d}{dt}\left(\frac{\partial V}{\partial \dot{x}}\right) = \dot{x}\frac{\partial}{\partial x}\left(\frac{\psi_x}{\eta_x \dot{x} + \eta_t}\right) + \frac{\partial}{\partial t}\left(\frac{\psi_x}{\eta_x \dot{x} + \eta_t}\right), \quad (21)$$

where,

$$\dot{x}\frac{\partial}{\partial x}\left(\frac{\psi_x}{\eta_x \dot{x} + \eta_t}\right) = \frac{\dot{x}\psi_{xx}}{\eta_x \dot{x} + \eta_t} - \frac{\dot{x}\psi_x(\eta_{xx}\dot{x} + \eta_{tx})}{(\eta_x \dot{x} + \eta_t)^2}$$

$$\frac{\partial}{\partial t}\left(\frac{\psi_x}{\eta_x \dot{x} + \eta_t}\right) = \frac{\psi_{xt}}{\eta_x \dot{x} + \eta_t} - \frac{\psi_x(\eta_{xt}\dot{x} + \eta_{tt})}{(\eta_x \dot{x} + \eta_t)^2}.$$

Further, the second term of the left-hand side of (19) is

$$\frac{\partial V}{\partial x} = \frac{\phi_{xx}\dot{x} + \phi_{xt}}{\eta_x \dot{x} + \eta_t} - \frac{V(\eta_{xx}\dot{x} + \eta_{xt})}{\eta_x \dot{x} + \eta_t} = \frac{\psi_{xx}\dot{x} + \psi_{xt}}{\eta_x \dot{x} + \eta_t}. \quad (22)$$

Finally, from (19), (21) and (22),

$$\dot{x}\frac{\partial}{\partial x}(\eta_x \dot{x} + \eta_t) + \frac{\partial}{\partial t}(\eta_x \dot{x} + \eta_t) = 0. \quad (23)$$

(23) can be rewritten as

$$\frac{d}{dt}\left(\frac{dT}{dt}\right) = 0. \quad (24)$$

with help of (3),

$$\frac{d}{dt}\left(\frac{dT}{dt}\frac{dX}{dT}\right) = \frac{d}{dt}\left(\frac{dX}{dt}\right) = 0. \quad (25)$$

From (24), (25)

$$\frac{d}{dt}\left(\frac{dX}{dt} - V\frac{dT}{dt}\right) = 0. \quad (26)$$

with help of (9),

$$\frac{\partial^2 \psi}{\partial t^2} = \dot{x}^2 \frac{\partial^2 \psi}{\partial x^2}. \quad (27)$$

Newly, the $\Psi$ satisfies the classical wave equation and vice versa. (13) certifies that the $\Psi$ defined as the solution of (27) always minimizes the Lagrangian function (14). However, (27) is merely the kinetic condition which the microscopic space-time should satisfy to consistent with (3) and (5). To consistent with the macroscopic motion equation, the governing equation that describes the fluctuation of the microscopic space-time should satisfy the following relationship between the kinetic energy and the momentum. It means that only the solution of (27) which satisfies the following relationship in addition, as a binding condition is the collect microscopic space-time.

$$E = \frac{P^2}{2m}, \quad (28)$$

where, m is mass of the particle. To make consideration of the binding condition (28), the following variable separation is applied to (27).

$$\psi = \alpha(x) \cdot \beta(t). \quad (29)$$

By substitution of (29) into (27), the following relationship as a general property of the wave equation is obtained.

$$\frac{\dot{x}^2}{\alpha(x)}\frac{d^2\alpha(x)}{dx^2} = \frac{1}{\beta(t)}\frac{d^2\beta(t)}{dt^2} = C. \quad (30)$$

As α(x) and β(t) express the magnitude of the fluctuation of the microscopic space-time, the divergent α(x) and β(t) are never acceptable physically, the constant C should be negative as follows,

$$C = -(2\pi\nu)^2, \quad (31)$$

where, $\nu$ is the frequency which is determined by (27). Thus,

$$\frac{d^2\alpha(x)}{dx^2} = -\left(\frac{2\pi\nu}{\dot{x}}\right)^2 \alpha(x), \quad (32)$$

$$\frac{d^2\beta(t)}{dt^2} = -(2\pi\nu)^2 \beta(t). \quad (33)$$

Multiplying (32) by β(t) and (33) by α(x), the following relationships are obtained,

$$\frac{\partial^2 \psi}{\partial x^2} = -\left(\frac{2\pi\nu}{\dot{x}}\right)^2 \psi = -\left(\frac{2\pi}{\lambda}\right)^2 \psi, \quad (34)$$

$$\frac{\partial^2 \psi}{\partial t^2} = -(2\pi\nu)^2 \psi. \quad (35)$$

Where, $\lambda$ is the wavelength. Here we assume that the frequency $\nu$ and the wavelength $\lambda$ which is defined by (34) and (35) respectively satisfy the following Einstein-de Broglie's formula,

$$E = h\nu, \quad P = \frac{h}{\lambda}. \quad (36)$$

where h is the Planck's constant. We substitute (34)-(36) into (28). As a result, the $\Psi$ satisfies the following wave equation,

$$\frac{\partial^2 \psi}{\partial t^2} = -\left(\frac{\hbar}{2m}\right)^2 \frac{\partial^4 \psi}{\partial x^4}. \quad (37)$$

where,

$$\hbar = \frac{h}{2\pi}. \quad (38)$$

And, as might be expected, the relationship of $\dot{x}$ =V/2 can be gotten. The (37) satisfies the binding condition (28) through the Einstein-de Broglie's formula, and is just the wave equation of the dispersive wave which is common of the vibration problem of span. It is known that in 1926 Schrödinger considered the equation (37) is the wave equation of particle motion by letters to Lorenz and Planck from Schrödinger [4]. As a general property, (37) is rewritten as,

$$\left(\frac{\partial}{\partial t} - i\frac{\hbar}{2m}\frac{\partial^2}{\partial x^2}\right)\left(\frac{\partial \psi}{\partial t} + i\frac{\hbar}{2m}\frac{\partial^2 \psi}{\partial x^2}\right) = 0,$$

or

$$\left(\frac{\partial}{\partial t} + i\frac{\hbar}{2m}\frac{\partial^2}{\partial x^2}\right)\left(\frac{\partial \psi}{\partial t} - i\frac{\hbar}{2m}\frac{\partial^2 \psi}{\partial x^2}\right) = 0.$$

Then, the $\Psi$ satisfy the following equations.

$$\frac{\partial \psi}{\partial t} + i\frac{\hbar}{2m}\frac{\partial^2 \psi}{\partial x^2} = 0, \quad (39)$$

$$\frac{\partial \psi}{\partial t} - i\frac{\hbar}{2m}\frac{\partial^2 \psi}{\partial x^2} = 0. \quad (40)$$

These are just the Schrödinger equation of $\Psi$ and $\Psi^*$ which is the complex conjugate function of $\Psi$. So, the general solution of (37) is given as a sum of the solutions of (39) and (40) as follows,

$$\psi + \psi^* = \text{Re}(\psi).$$

This means the solution $\Psi$ is always the real function and is not the general solution of the Schrödinger equation that may describe by the complex function. Further, as (37) includes the second-order temporal differentiation, on the standard probabilistic interpretation (37) never certify the conservation law of the probability density. So, we can guess that Schrödinger abandoned (37) as the correct wave equation of particle motion. On this paper, (37) is re-evaluated using Dirac's formula [5] that was used to the relativistic wave equation for the electron from the Klein-Gordon equation. First, we extend (37) to the three dimensional one as follows,

$$\frac{\partial^2 \psi}{\partial t^2} = -\left(\frac{\hbar}{2m}\right)^2 \left(\frac{\partial^4 \psi}{\partial x^4} + \frac{\partial^4 \psi}{\partial y^4} + \frac{\partial^4 \psi}{\partial z^4}\right). \quad (41)$$

To reduce the order of the temporal differentiation, (41) can be rewritten as follows

$$\hbar \frac{\partial \psi}{\partial t} = \pm i \frac{\hbar^2}{2m}\left(\sigma_x \frac{\partial^2 \psi}{\partial x^2} + \sigma_y \frac{\partial^2 \psi}{\partial y^2} + \sigma_z \frac{\partial^2 \psi}{\partial z^2}\right). \quad (42)$$

This is the reasonable modification of (39) and (40). To duplicate (41) by multiplying (42) by the differential operator of both sides, the following conditions should be satisfied.

$$\sigma_i \sigma_i = 1, \quad \sigma_i \sigma_j + \sigma_j \sigma_i = 0 \quad (i \neq j). \quad (43)$$

If the $\sigma_x$, $\sigma_y$, $\sigma_z$ are the following scheme of the Pauli matrices, then (42) can be duplicated to (41).

$$\sigma_x = \begin{pmatrix} 0 & 1 \\ 1 & 0 \end{pmatrix}, \quad \sigma_y = \begin{pmatrix} 0 & -i \\ i & 0 \end{pmatrix}, \quad \sigma_z = \begin{pmatrix} 1 & 0 \\ 0 & -1 \end{pmatrix}. \quad (44)$$

Then, the $\Psi$ should be written as a vector, and (42) can be rewritten as follows,

$$\hbar \frac{\partial}{\partial t}\begin{pmatrix} \psi_1 \\ \psi_2 \end{pmatrix} = i \frac{\hbar^2}{2m} \frac{\partial^2}{\partial x^2}\begin{pmatrix} 0 & 1 \\ 1 & 0 \end{pmatrix}\begin{pmatrix} \psi_1 \\ \psi_2 \end{pmatrix}$$

$$+ i\frac{\hbar^2}{2m}\frac{\partial^2}{\partial y^2}\begin{pmatrix} 0 & -i \\ i & 0 \end{pmatrix}\begin{pmatrix} \psi_1 \\ \psi_2 \end{pmatrix} + i\frac{\hbar^2}{2m}\frac{\partial^2}{\partial z^2}\begin{pmatrix} 1 & 0 \\ 0 & -1 \end{pmatrix}\begin{pmatrix} \psi_1 \\ \psi_2 \end{pmatrix}, \quad (45)$$

$$\hbar \frac{\partial}{\partial t}\begin{pmatrix} \psi_3 \\ \psi_4 \end{pmatrix} = -i\frac{\hbar^2}{2m}\frac{\partial^2}{\partial x^2}\begin{pmatrix} 0 & 1 \\ 1 & 0 \end{pmatrix}\begin{pmatrix} \psi_3 \\ \psi_4 \end{pmatrix}$$

$$- i\frac{\hbar^2}{2m}\frac{\partial^2}{\partial y^2}\begin{pmatrix} 0 & -i \\ i & 0 \end{pmatrix}\begin{pmatrix} \psi_3 \\ \psi_4 \end{pmatrix} - i\frac{\hbar^2}{2m}\frac{\partial^2}{\partial z^2}\begin{pmatrix} 1 & 0 \\ 0 & -1 \end{pmatrix}\begin{pmatrix} \psi_3 \\ \psi_4 \end{pmatrix}. \quad (46)$$

As this paper is considered under the nonrelativistic approximation, the $\Psi$ is the two dimensional vector unlike the Dirac equation. (45), (46) can be rewritten with each vector component as follows,

$$\hbar \frac{\partial \psi_1}{\partial t} = i\frac{\hbar^2}{2m}\frac{\partial^2 \psi_2}{\partial x^2} + \frac{\hbar^2}{2m}\frac{\partial^2 \psi_2}{\partial y^2} + i\frac{\hbar^2}{2m}\frac{\partial^2 \psi_1}{\partial z^2}, \quad (47)$$

$$\hbar \frac{\partial \psi_2}{\partial t} = i\frac{\hbar^2}{2m}\frac{\partial^2 \psi_1}{\partial x^2} - \frac{\hbar^2}{2m}\frac{\partial^2 \psi_1}{\partial y^2} - i\frac{\hbar^2}{2m}\frac{\partial^2 \psi_2}{\partial z^2}, \quad (48)$$

$$\hbar \frac{\partial \psi_3}{\partial t} = -i\frac{\hbar^2}{2m}\frac{\partial^2 \psi_4}{\partial x^2} - \frac{\hbar^2}{2m}\frac{\partial^2 \psi_4}{\partial y^2} - i\frac{\hbar^2}{2m}\frac{\partial^2 \psi_3}{\partial z^2}, \quad (49)$$

$$\hbar \frac{\partial \psi_4}{\partial t} = -i\frac{\hbar^2}{2m}\frac{\partial^2 \psi_3}{\partial x^2} + \frac{\hbar^2}{2m}\frac{\partial^2 \psi_3}{\partial y^2} + i\frac{\hbar^2}{2m}\frac{\partial^2 \psi_4}{\partial z^2}. \quad (50)$$

(47)-(50) are considered as the Schrödinger equations that include the spin angular momentum. To confirm it, the free particle motion along the z-direction is considered by (45). Then, (45) is simplified as

$$\hbar \frac{\partial}{\partial t}\begin{pmatrix} 1 & 0 \\ 0 & 1 \end{pmatrix}\begin{pmatrix} \psi_1 \\ \psi_2 \end{pmatrix} = i\frac{\hbar^2}{2m}\frac{\partial^2}{\partial z^2}\begin{pmatrix} 1 & 0 \\ 0 & -1 \end{pmatrix}\begin{pmatrix} \psi_1 \\ \psi_2 \end{pmatrix}. \quad (51)$$

As a solution of ($\Psi_1$, $\Psi_2$), we have the following harmonic one,

$$\begin{pmatrix} \psi_1 \\ \psi_2 \end{pmatrix} = \begin{pmatrix} u_1 \\ u_2 \end{pmatrix} e^{i\hbar^{-1}(Pz - Et)}. \quad (52)$$

By substitution of (52) into (51),

$$-iE\begin{pmatrix} 1 & 0 \\ 0 & 1 \end{pmatrix}\begin{pmatrix} u_1 \\ u_2 \end{pmatrix} = -i\frac{P^2}{2m}\begin{pmatrix} 1 & 0 \\ 0 & -1 \end{pmatrix}\begin{pmatrix} u_1 \\ u_2 \end{pmatrix}, \quad (53)$$

$$\begin{pmatrix} E - \frac{P^2}{2m} & 0 \\ 0 & E + \frac{P^2}{2m} \end{pmatrix}\begin{pmatrix} u_1 \\ u_2 \end{pmatrix} = 0. \quad (54)$$

As ($u_1$, $u_2$) should have non zero solution, then

$$\left(E - \frac{P^2}{2m}\right)\left(E + \frac{P^2}{2m}\right) = 0. \quad (55)$$

Under the nonrelativistic approximation, the kinetic energy should be positive. Then,

$$E = \frac{P^2}{2m}. \quad (56)$$

This means (54) satisfies the correct relationship between the kinetic energy and the momentum as an eigenvalue. From the eigenvalue (56), the eigenvector (52) is written as follows,

$$\begin{pmatrix} \psi_1 \\ \psi_2 \end{pmatrix} = \begin{pmatrix} u_1 \\ 0 \end{pmatrix} e^{i\hbar^{-1}(Pz-Et)} = u_1 \begin{pmatrix} 1 \\ 0 \end{pmatrix} e^{i\hbar^{-1}(Pz-Et)}. \quad (57)$$

Similarly, we consider the free particle motion along the z direction by (46). Then, (46) is

$$\hbar \frac{\partial}{\partial t}\begin{pmatrix} 1 & 0 \\ 0 & 1 \end{pmatrix}\begin{pmatrix} \psi_3 \\ \psi_4 \end{pmatrix} = i\frac{\hbar^2}{2m}\frac{\partial^2}{\partial z^2}\begin{pmatrix} -1 & 0 \\ 0 & 1 \end{pmatrix}\begin{pmatrix} \psi_3 \\ \psi_4 \end{pmatrix}. \quad (58)$$

Under the harmonic solution of ($\Psi_3$, $\Psi_4$), (58) is

$$\begin{pmatrix} E + \frac{P^2}{2m} & 0 \\ 0 & E - \frac{P^2}{2m} \end{pmatrix}\begin{pmatrix} u_3 \\ u_4 \end{pmatrix} = 0. \quad (59)$$

Thus, the eigenvector ($\Psi_3$, $\Psi_4$) is

$$\begin{pmatrix} \psi_3 \\ \psi_4 \end{pmatrix} = \begin{pmatrix} 0 \\ u_4 \end{pmatrix} e^{i\hbar^{-1}(Pz-Et)} = u_4 \begin{pmatrix} 0 \\ 1 \end{pmatrix} e^{i\hbar^{-1}(Pz-Et)}. \quad (60)$$

(57) and (60) can be written as follows,

$$\psi_\uparrow = u_\uparrow \begin{pmatrix} 1 \\ 0 \end{pmatrix} e^{i\hbar^{-1}(Pz-Et)}, \quad (61)$$

$$\psi_\downarrow = u_\downarrow \begin{pmatrix} 0 \\ 1 \end{pmatrix} e^{i\hbar^{-1}(Pz-Et)}. \tag{62}$$

These are just the wave functions that can describe the spin up, and the spin down. Actually, using the following Pauli's spin operator,

$$S_z = \frac{\hbar}{2}\begin{pmatrix} 1 & 0 \\ 0 & -1 \end{pmatrix}, \tag{63}$$

then

$$S_z\psi_\uparrow = \frac{\hbar}{2}u_\uparrow \begin{pmatrix} 1 & 0 \\ 0 & -1 \end{pmatrix}\begin{pmatrix} 1 \\ 0 \end{pmatrix} e^{i\hbar^{-1}(Pz-Et)} = \frac{\hbar}{2}\psi_\uparrow, \tag{64}$$

$$S_z\psi_\downarrow = \frac{\hbar}{2}u_\downarrow \begin{pmatrix} 1 & 0 \\ 0 & -1 \end{pmatrix}\begin{pmatrix} 0 \\ 1 \end{pmatrix} e^{i\hbar^{-1}(Pz-Et)} = -\frac{\hbar}{2}\psi_\downarrow. \tag{65}$$

These describe the wave function with the spin angular momentum correctly. The reasoning of the fluctuated space-time in the microscopic systems, can lead to the Schrödinger equation with the spin angular momentum naturally.

In addition, if (37) is separated into (39) and (40), the general solution is the real function. On the other hands, using (45) and (46) that is separated by the Pauli matrices, the another general solution of (41) is as follows,

$$\psi_\uparrow + \psi_\downarrow = \begin{pmatrix} u_\uparrow \\ u_\downarrow \end{pmatrix} e^{i\hbar^{-1}(Pz-Et)}. \tag{66}$$

This solution is not the real function and describes the superimposed state of the spin up and the spin down. Actually, the inner product of (66),

$$|\psi_\uparrow + \psi_\downarrow|^2 = u_\uparrow \cdot u_\uparrow^* + u_\downarrow \cdot u_\downarrow^*, \tag{67}$$

indicates the superposition of the spin up and the spin down.

§3. Discussions

Firstly, let us consider the meaning of the $\Psi$. The $\Psi$ is defined by (9) as follows,

$$\psi = \phi - V\eta = X - V \cdot T. \tag{68}$$

By the modification, (68) is rewritten as

$$X = V \cdot T + \psi. \tag{69}$$

In the macroscopic dynamic law of the free particle, the $\Psi$ is constant that corresponds to the initial condition. On the other hand, because of the existence of the fluctuations of the microscopic space-time by the particle motion in the microscopic motion law, the $\Psi$ never becomes constant. That is, the $\Psi$ can be interpreted as the *'disparity'* or *'deviation'* between the Newton's law and the microscopic dynamic law. Then, with the limiting behavior of $\Psi \to 0$, (69) can be completely corresponding to the Newton's law whose initial position is assumed to be 0.

When the fluctuations of the microscopic space-time are negligible small, the microscopic space-time systems may be equivalent to the macroscopic ones. Then, it is extremely natural that the microscopic dynamical law can be close to the Newton's law defined on the macroscopic systems. The present concept of the microscopic dynamic law says that the wave function $\Psi$ and the macroscopic i.e. Newton's one can be related directly without using the Ehrenfest theory. The $\Psi$ defined by (9) has the dimension of distance. If the $\Psi$ is the complex function, the inner product of the $\Psi$ and $\Psi^*$, that is the complex conjugate function of $\Psi$, has the dimension of the distance squared. When the inner product of ($\Psi \cdot \Psi^*$) is normalized to unity, the standard probability interpretation of the Copenhagen school can be applicable to understand the meaning of the $\Psi$.

On two particle systems, the wave function is defined as a six dimensional 'wave' that includes the coordinates of particle A and particle B. Even on this case, the $\Psi$ can be considered that it express the *'disparity'* with the Newton's law. The disparity is caused by the fluctuations of the microscopic space-time through the motion of the particle A and B. So, only the $\Psi$ of the one particle system is wavelike exactly.

Secondarily, let us consider the difference between the macroscopic and microscopic motion of the particle. If the microscopic space-time system has the fluctuation for the macroscopic and Euclidean system where we can recognize and define the particle motion, we should accept the particle existence at the *'past'* and the *'future'* of the microscopic time at the *'macroscopic present'*. Then, the particle may not have the *'deterministic'* path at the macroscopic 'present', as the macroscopic 'present' includes the microscopic 'past' and 'future'. The relativity of the 'present' between the macroscopic and microscopic times clarifies the deference between the quantum mechanics and the classical ones. That is, the classical dynamic equation is determined along the unidirectional time axis from the past to the future. But, in the microscopic systems, the 'past' and 'future' of microscopic time at the macroscopic 'present' disturb to predict the *'deterministic'* particle motion as the meaning of the macroscopic systems.

In that sense, our actions to observe the physical quantity of the particle motion may make the 'present' of the particle on the macroscopic systems. The observation is the frozen figure of the microscopic motion at the macroscopic 'present'. It means that the observation is the action to hypostatize the particle motion at the macroscopic 'present'. Then, the observation gives an unavoidable interaction to the particle motion. As well as the interaction is usually estimated by the Compton effects on the Heisenberg uncertainty principle, it may be considered as the interference between the double fluctuated space-time due to the particle motion and the observation.

From here onwards, even if the fluctuation of the microscopic space-time leads to the wave nature of the particle, the probability interpretation can have applicability to predict the particle motion as above mentioned. However, the most important difference with the probability interpretation is that the particle is just the 'particle' and the wave nature is caused by the fluctuation of the microscopic space-time systems on this paper. That is, every particle has and recognizes their own deterministic paths in the microscopic space-time systems. However, on the macroscopic systems, due to 'the relativity of the present' and the unavoidable interaction, their microscopic paths can never be recognized and observed.

The particle can choose one path from the infinite options of the paths consistent with the magnitude of fluctuation of microscopic space-time. But, it is impossible to predict which path from the potential paths is hypostatized, as the initial conditions, which are unpredictable in the macroscopic space-time, change the state of the fluctuations. And then, along the hypostatized path which satisfies the deterministic equations (45), (46), the particle moves in the microscopic space-time. On the case of one particle system, the observation to confirm which path is hypostatized gives the unavoidable interaction to the path. While on the other hand, suppose the case of two particles system, for instant, the splitting of one particle to two particles A and B. Both particles shall move a long distance in the counter direction after the splitting. After adequate time, if the physical quantity of either particle A is observed, the quantity of the other particle B is predicted exactly. Assuming that both particles confirm one hypostatized path that satisfies the deterministic equations (45), (46) at the instant of the splitting, it is quite reasonable that the observation of the state of the either particle A determines the state of the other particle B. In the present case, the action at a distance between the observation of state of the particle A and the fixedness of the state of the particle B is never required. Considering the particles can hypostatize one path from the infinite paths and move along the path, the ERP [6] paradox is nonexistent. The concept of the existence of infinite paths of the particle motion is equivalent to the quantum mechanics based on the path integral by Feynman [7]. This paper advocates that the infinite paths are caused by the fluctuation on microscopic space-time. On that point, setting the reduction of wave pocket aside, this paper never disallows the standard interpretation of the quantum mechanics, but supplements the cause of the difference with the classical mechanics and the quantum mechanics.

If the problem of the two-slit experiment is examined using the present interpretation, the particle never passes the two-slit, and should pass either one of the two. Then, the interaction between the fluctuated microscopic space-time due to the particle motion and due to the potential of two-slit may make the interference of the fluctuations. The interference which limit the path of particle, cause the fringe pattern of particle position on the screen. Then, there is no repugnance between the existence microscopic path and the figuration of the fringe pattern. The reduction of wave packet may be eliminated naturally.

This interpretation that advocates the existence of the microscopic path resembles to the Nelson's theory [8]. It also can derive the Schrödinger equation using stochastic process. On the process, Nelson required the fluctuation term to the Newton's law due to the Brownian motions and the new local differential to define the velocity on the zigzag path. And the other, as the present concept never requires these assumptions, it may be simpler.

§4. Conclusion

On the standard quantum mechanics, the wave nature of the particle is considered as the attribution of particle itself. And the other, on this paper, the wave nature is the attribution of the microscopic space-time. The particle motion gives fluctuations to the microscopic space-time, and vice versa, the space-time determined the particle motion. Then, on the macroscopic and Euclidean space-time systems, the particle motion may be recognized as the wave nature. Using such a concept, the wave-particle duality is understood naturally. Further, the Schrödinger equation that includes the spin angular momentum is derived naturally. Even this concept, the probability interpretation can be applicable to predict the physical quantity. But, as the present concept never requires the reduction of wave pocket on the observation, this is looked more rational than the concept based on the wave nature of particle.